\documentclass[aps,prd,10pt,twocolumn,nofootinbib,floatfix,showkeys]{revtex4-2}

\usepackage{graphicx}
\usepackage{amsmath,amssymb}
\usepackage{booktabs}
\usepackage[hidelinks]{hyperref}
\usepackage{url}
\newcommand{\mango}{\textsc{mango}}

\newcommand{\costheta}{\cos\theta_z}
\newcommand{\dmsq}{\Delta m^2_{31}}

\newcommand{\JPG}{J. Phys. G}
\newcommand{\CPC}{Comput. Phys. Commun.}
\newcommand{\PR}{Phys. Rev.}
\newcommand{\etal}{\textit{et al.}}

\begin{document}

\title{\textbf{MANGO}: An Autodiff Neutrino Oscillation Engine\\[0.15em]
for Differentiable Analysis Pipelines}

\author{P.~Granger}
\thanks{ORCID: \href{https://orcid.org/0000-0002-8333-4393}{0000-0002-8333-4393}}
\email{pierre.granger@cern.ch}
\affiliation{CERN, CH-1211 Geneva 23, Switzerland}

\date{\today}

\begin{abstract}
Computing neutrino oscillation probabilities is a solved problem; computing
their derivatives is not. We present \mango{} (\textbf{M}ANGO: \textbf{A} \textbf{N}eutrino \textbf{G}radient \textbf{O}scillator), a composable oscillation engine
in which every computed quantity is differentiable with respect to all inputs,
including propagation geometry, detector depth, and individual Earth-shell
densities. None of these quantities appear in traditional analytic probability
formulas. The engine supports vacuum, constant-density, layered-PREM,
arbitrary-profile, and adiabatic solar propagation, alongside front-ends for
non-standard interactions, $3{+}N$ sterile states, decoherence, and
non-unitary mixing. Because reverse-mode cost depends on output rather than
input dimension, evaluating sensitivities incurs a constant $2.5$--$3\times$
forward-pass overhead. As a result, calculating sensitivities for all $369$
density, electron-fraction, and shell-radius parameters of a layered Earth
costs no more than the $6$ standard oscillation parameters.
By maintaining exact sensitivity signals, \mango{} allows gradients to flow continuously past the probability
stage and through detector response, event weighting, binning, and
likelihoods. We demonstrate this on a stylized Earth-tomography analysis. In a
single pass, \mango{} computes the marginalized uncertainty on a six-zone
radial density model and, by differentiating through the inverse Fisher
matrix, evaluates its sensitivity with respect to detector angular resolution.
This provides an experimental-design metric unreachable using traditional
analytic probability formulas. Three-flavor and layered-Earth probabilities
match external benchmarks (\textsc{OscProb}, \textsc{NuFast-Earth}) to within
$10^{-9}$--$10^{-5}$, while all forward models, BSM limits, and differentiation paths
are verified against exact analytic solutions and finite differences.
\end{abstract}


\maketitle

\section{Introduction}
\label{sec:intro}

Although modern software reliably calculates neutrino oscillation probabilities across diverse propagation regimes, evaluating their exact derivatives and embedding them into end-to-end differentiable pipelines remains an unaddressed challenge.
A rich ecosystem of mature, highly optimized packages exists, including \textsc{Prob3++}~\cite{prob3}, \textsc{OscProb}~\cite{oscprob}, \textsc{GLoBES}~\cite{globes}, \textsc{nuSQuIDS}~\cite{nusquids}, and most recently \textsc{NuFast}~\cite{nufast}.
While these engines excel at forward probability evaluations, none provide exact, automated derivatives, nor can they be embedded into downstream differentiable computation chains.

This architectural limitation is now a practical bottleneck.
Modern analysis techniques, ranging from Hamiltonian Monte Carlo (HMC) and variational inference to neural surrogates and gradient-based experimental design, require derivatives through the entire analysis pipeline, extending from oscillation probabilities and flux models down to detector response, histogram binning, and final test statistics.
What has been missing is a composable, differentiable oscillation engine to serve as the upstream foundation for such chains.

Differentiable programming has already transformed high-energy physics through differentiable inference and analysis/detector co-optimization~\cite{pyhf,neos,e2ewhitepaper,inferno,tomopt}, as well as end-to-end differentiable detector simulation~\cite{gasiorowski}.
Similar paradigms underpin differentiable Fisher-matrix estimation in cosmology~\cite{jaxcosmo}, gravitational-wave astronomy~\cite{gwfast}, and optical instrument design~\cite{dlux}.
Within neutrino physics, the value of exact probability derivatives was recently recognized: \textsc{CHIC}~\cite{chic} derives closed-form Cayley--Hamilton expressions for three-flavor constant-density probabilities and their derivatives with respect to the Hamiltonian parameters, and demonstrates their analysis value through ``oscillograd'' visualizations.
While analytic derivatives are fast and exact, they must be derived and implemented by hand for a specific propagation regime; \textsc{CHIC} is consequently restricted to three-flavor constant-density propagation and lists varying-density media, beyond-standard-model effects, and sterile states as future work.
In parallel, automatic differentiation has begun powering differentiable global likelihood frameworks in Julia (\textsc{Newtrinos.jl}~\cite{newtrinos}).

\mango{} (\textbf{M}ANGO: \textbf{A} \textbf{N}eutrino \textbf{G}radient \textbf{O}scillator) takes a complementary, automatic differentiation (AD) approach implemented natively in JAX~\cite{jax}.
Rather than deriving analytic derivatives for a specific case, it implements a general propagation core and obtains derivatives by automatic differentiation.
The consequences are threefold:

\begin{itemize}
    \item \textbf{Coverage:} Derivatives come for free for every configuration
    the code supports, including the layered PREM Earth, continuous solar profiles with adiabatic MSW propagation, arbitrary profiles, NSI,
    and sterile sectors.
    \item \textbf{Reach:} One can differentiate with respect to quantities that
    are not parameters of any analytic probability formula. The zenith angle,
    the atmospheric production height, and the depth of the detector all enter
    through the chord geometry; the per-shell densities, electron fractions, and
    boundary radii of the Earth model together form precisely the
    probability-space Jacobian that oscillation-based Earth tomography starts
    from~\cite{tomowinter,tomokelly}.\footnote{This is distinct from absorption-based tomography of TeV neutrinos~\cite{tomodonini}, which relies on high-energy attenuation rather than oscillation phases, requiring a transport treatment that changes total neutrino number (section~\ref{sec:limits}).}
    \item \textbf{Composability:} Most importantly, the AD graph does not
    terminate at the probability calculation. Weights, detector response,
    histogram binning, and the resulting test statistic can all be JAX
    functions, so a single \texttt{jax.grad} call returns the gradient of a
    physics-analysis figure of merit with respect to any upstream quantity.
    Section~\ref{sec:worked} carries this through in full for an
    Earth-tomography measurement, illustrating the broader paradigm where
    physics models, instrument design parameters, and analysis pipelines can be
    jointly differentiated and optimized end to end.
\end{itemize}

On standard axes \mango{} matches existing tools by supporting arbitrary radial matter profiles and $3{+}N$ flavors (as provided by \textsc{OscProb}~\cite{oscprob} and \textsc{nuSQuIDS}~\cite{nusquids}) alongside GPU execution (as provided by \textsc{CUDAProb3}~\cite{cudaprob3}).
This initial release focuses on establishing a fully differentiable engine for core oscillation physics; extending coverage to full density-matrix transport (as in \textsc{nuSQuIDS}), complete experiment simulation (as in \textsc{GLoBES}~\cite{globes}), or exhaustive BSM suites (as in \textsc{OscProb}) remains for future iterations.
While \textsc{CHIC}~\cite{chic} provides derivatives exclusively for constant-density propagation, none of these tools supply exact, input-agnostic derivatives that seamlessly compose into downstream analysis pipelines.
That is the primary focus of this work.

This paper documents the physics formalism and extended Hamiltonians (section~\ref{sec:formalism}), the numerical backends (section~\ref{sec:numerics}), the differentiability model and its cost (section~\ref{sec:autodiff}), the validation programme (section~\ref{sec:validation}), and a worked Earth-tomography example carrying derivatives through a Fisher matrix and its inverse (section~\ref{sec:worked}).

\section{Formalism}
\label{sec:formalism}

\subsection{Standard three-flavor propagation and Earth geometry}

In the flavor basis, propagation through a medium with electron number density $N_e(x)$ is governed by \begin{equation} \begin{split} H(x) = {}& \frac{1}{2E}\,U\,\mathrm{diag}\!\left(0,\Delta m^2_{21},\dmsq\right)U^\dagger \\ & + \mathrm{diag}\!\left(V_{\rm CC}(x),0,0\right), \end{split} \label{eq:hamiltonian} \end{equation} where $V_{\rm CC}(x)=\sqrt{2}\,G_F N_e(x)$, $G_F$ is the Fermi coupling constant, and $U$ is the PMNS matrix in the PDG parametrization $R_{23}U_{13}(\delta)R_{12}$.
Mass splittings are referred to $m_1$, so that the ordering is set by the sign of $\dmsq$: normal for $\dmsq>0$, inverted for $\dmsq<0$, with no other change of convention.
Every calculation in this paper uses the NuFIT 5.1 normal-ordering best fit without SK atmospheric data~\cite{nufit}.
Antineutrinos are obtained by $U\to U^*$, with every matter potential flipping sign ($V_{\rm CC}\to -V_{\rm CC}$, and similarly $V_{\rm NC}\to -V_{\rm NC}$ for neutral-current terms).

\begin{figure}[!t]
\centering
\includegraphics[width=\columnwidth]{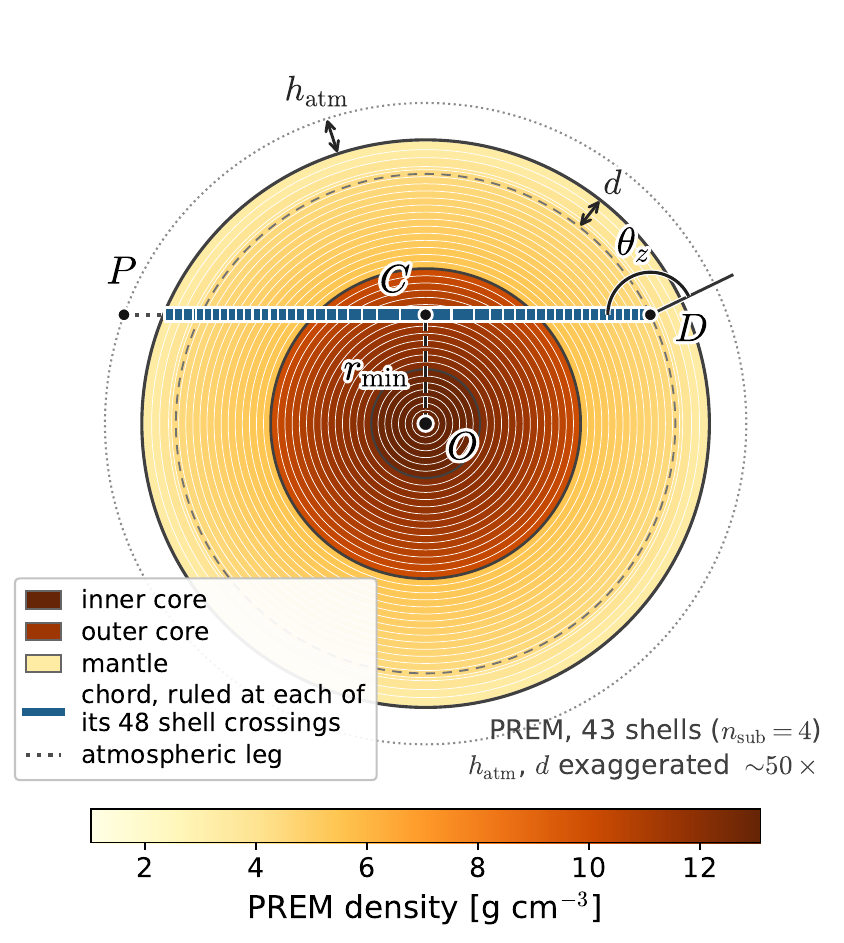}
\caption{The chord construction of equation~(\ref{eq:chord}). A neutrino
produced at $P$ travels in a straight line to a detector at $D$, passing
closest to the Earth's centre $O$ at $C$, a distance $r_{\min}$ away. Each leg
is evaluated in closed form and shell boundaries are crossed exactly, so the
segment lengths, and hence the probability, are differentiable functions of
$\costheta$, $h_{\rm atm}$ and $d$ even though none of the three appears in
the Hamiltonian. The shells and their colours are the actual default table,
$43$ shells at $n_{\rm sub}=4$, rather than a three-zone cartoon: the
trajectory drawn ($\costheta=-0.9$, crossing the outer core) is cut into the
$48$ constant-density segments the white rules mark out, and the ordered
product of equation~(\ref{eq:product}) runs over those. Those rules are the
shell crossings themselves, not a periodic pattern, which is why their
spacing varies by a factor of a few hundred: with $x=\sqrt{r^2-r_{\min}^2}$
along the chord the spacing $\mathrm{d}x/\mathrm{d}r$ diverges as $r\to
r_{\min}$, so near $C$ the trajectory runs almost parallel to the boundaries
and stays inside a single shell for a long path, while near the surface the
thin crustal shells are crossed one after another. Only $h_{\rm atm}$ and $d$
are exaggerated, by about $50\times$, since at true scale neither would be a
visible fraction of an Earth radius.}
\label{fig:chordgeom}
\end{figure}

The Earth is described by PREM, the Preliminary Reference Earth Model~\cite{prem}, as a table of spherical shells with configurable sub-shell refinement and a two-zone electron fraction $Y_{e}$, the number of electrons per nucleon, which converts the mass density of shell $k$ into the electron number density of equation~(\ref{eq:hamiltonian}) via $N_{e,k}=Y_{e,k}\,\rho_k/m_N$, where $m_N\equiv1\,\mathrm{u}=1/N_A$ is the atomic mass unit with $N_A$ Avogadro's number (the code evaluates $Y_e\rho N_A$ directly).
For a detector at depth $d$ and production at altitude $h_{\rm atm}$, the trajectory chord is parametrized by the distance from closest approach (figure~\ref{fig:chordgeom}): \begin{equation} \begin{gathered} s(r)=\sqrt{r^2-r_{\min}^2}\ ,\\ r_{\min}=(R_\oplus-d)\sqrt{1-\costheta^2}\ , \end{gathered} \label{eq:chord} \end{equation} where $\theta_z$ is the zenith angle and $R_\oplus$ is the Earth's radius.

This construction underpins the differentiability of the framework: $\costheta$, $h_{\rm atm}$ and $d$ determine the physical segment lengths $L_k$ through equation~(\ref{eq:chord}) and its shell intersections, rather than appearing inside the Hamiltonian of equation~(\ref{eq:hamiltonian}).
Because $s(r)$ is differentiable in closed form, derivatives such as $\partial P/\partial\costheta$, $\partial P/\partial h_{\rm atm}$ and $\partial P/\partial d$ exist and are exact (section~\ref{sec:autodiff}).
The segment lengths are smooth functions of $\costheta$ everywhere except on a measure-zero set of angles at which the chord grazes a shell boundary and the entering segment length acquires a square-root cusp; those critical angles are enumerable and discussed in section~\ref{sec:limits}.

For a piecewise-constant profile the evolution operator is the ordered product of segment propagators, \begin{equation} S = S_N S_{N-1}\cdots S_1, \qquad S_k=\exp(-iH_k L_k), \label{eq:product} \end{equation} and $P_{\alpha\beta}=|S_{\beta\alpha}|^2$.
The product is accumulated with \texttt{lax.scan} over a fixed-length segment list of $2N_{\rm shell}+1$ entries, where $N_{\rm shell}$ is set once by the shell table and not by the trajectory: the refinement parameter $n_{\rm sub}$ fixes a target radial step $R_\oplus/(10\,n_{\rm sub})$, allocating sub-shells in proportion to PREM region thicknesses.
Segment list lengths are deliberately \emph{not} adaptive in $\costheta$: shells the chord never reaches receive zero segment length ($L_k=0$) and act as identity propagators ($S_k=I$).
This trades a small amount of redundant computation at shallow angles for a fixed computational graph that \texttt{jit} and \texttt{vmap} compile once and reuse across entire zenith batches.

All internal arithmetic is performed in natural units (energies in eV, lengths in eV$^{-1}$) in IEEE double precision (\texttt{float64}), which \mango{} enables on import; unit conversions occur only at the API boundary.

\begin{figure*}[!t]
\includegraphics[width=\textwidth]{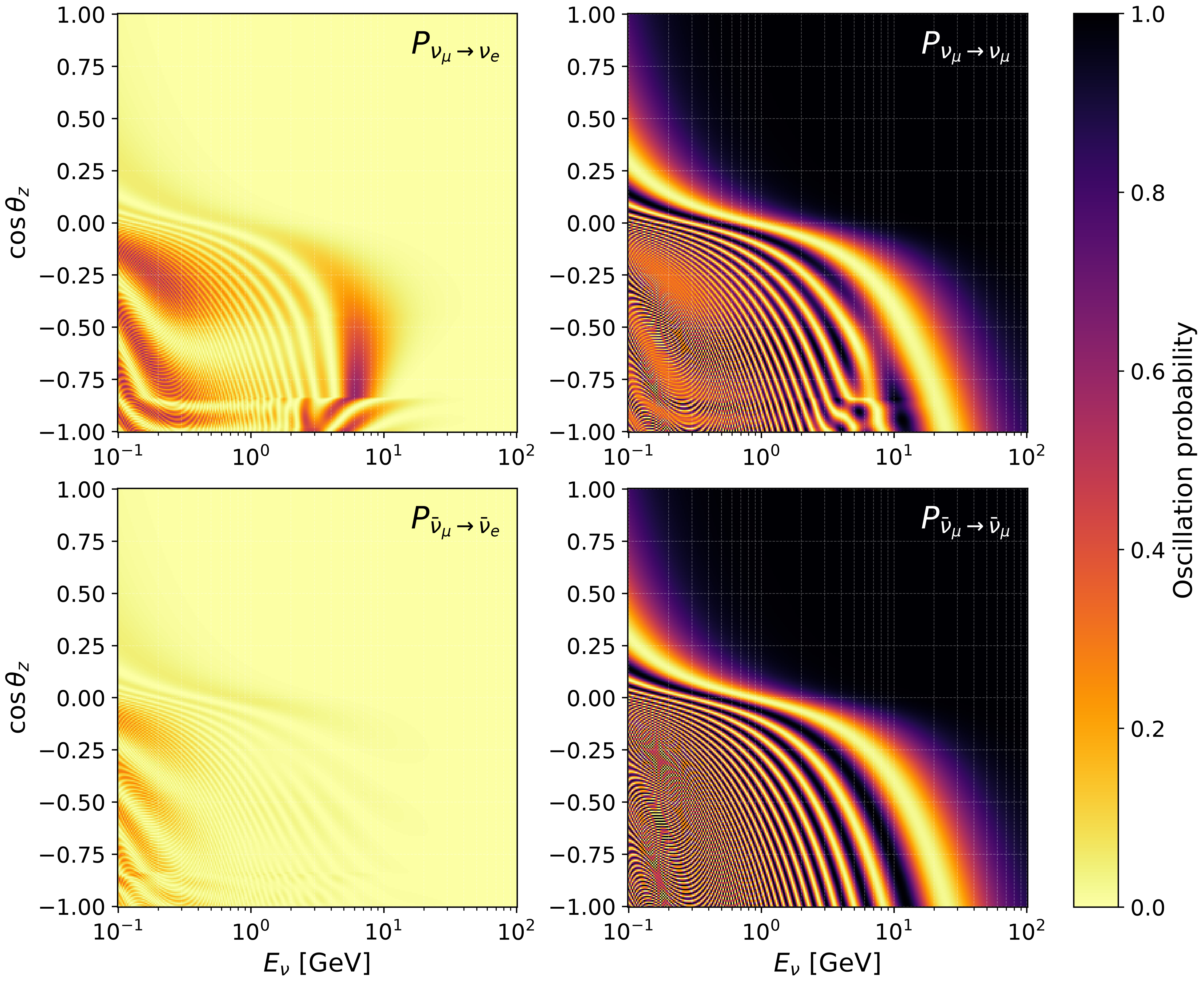}
\caption{Atmospheric oscillograms computed with \mango{} through the layered
PREM Earth: $\nu$ and $\bar\nu$ channels $\nu_\mu\to\nu_e$ and
$\nu_\mu\to\nu_\mu$ as a function of energy and $\costheta$. The MSW
resonance, the neutrino/antineutrino matter asymmetry, the core--mantle
boundary at $\costheta\simeq-0.84$ and the dense near-horizon oscillation
structure are all reproduced. Computed at the NuFIT 5.1 point of
section~\ref{sec:formalism} on a $300\times300$ grid,
$E_\nu\in[0.1,100]\,$GeV logarithmically and $\costheta\in[-1,1]$, with
sub-shell refinement $n_{\rm sub}=3$, production height $h_{\rm atm}=15\,$km,
zero detector depth and the default two-zone electron fraction.}
\label{fig:oscillogram}
\end{figure*}

Figure~\ref{fig:oscillogram} shows the atmospheric oscillograms this machinery produces; they are discussed as a qualitative check in section~\ref{sec:validation}.

\subsection{Beyond-standard-model Hamiltonians and extended sectors}

The propagation core generalizes to an arbitrary number of flavors $N$, accepting a general mixing matrix $U \in \mathbb{C}^{N\times N}$ and mass spectrum.
This allows extended sectors to be constructed via parameter front-ends without modifying internal propagation routines.

\paragraph{Non-standard interactions.} Vector matter NSI modify the matter
potential on the active $3\times3$ block,
\begin{equation}
H_{\rm matter} = V_{\rm CC}\left(\mathrm{diag}(1,0,0) + \varepsilon\right),
\qquad \varepsilon^\dagger = \varepsilon ,
\label{eq:nsi}
\end{equation}
where the six independent entries $\varepsilon_{ee},\varepsilon_{\mu\mu},
\varepsilon_{\tau\tau}$ (real) and $\varepsilon_{e\mu},\varepsilon_{e\tau},
\varepsilon_{\mu\tau}$ (complex) are treated as ordinary differentiable
parameters. Note the normalization: $\varepsilon$ is defined relative to
$V_{\rm CC}=\sqrt{2} G_{\mathrm{F}} N_e$, matching the standard propagation-NSI
convention shared by \textsc{OscProb} and \textsc{nuSQuIDS}. It differs by known
factors from the quark-level $\varepsilon^{u,d}$ conventions used in global fits,
and the two must not be compared directly without an explicit $Y_e/Y_n$
translation. For antineutrinos, $\varepsilon\to\varepsilon^{*}$ alongside
$U\to U^{*}$ and $V\to -V$.

\paragraph{Sterile neutrinos.} $3{+}N_s$ models are constructed by supplying
an enlarged mixing matrix and mass spectrum, as explored in global searches ranging from additional light species to extra-dimensional bulk fields~\cite{ettengruber_species,eller_bulk}. Sterile states carry no electroweak
charged- or neutral-current couplings; after removing the flavor-universal
active neutral-current term, the sterile diagonal of the matter Hamiltonian
receives the relative shift
\begin{equation}
V_{\rm NC} = \tfrac{1}{2}\sqrt{2}\,G_{\mathrm{F}} N_n ,\qquad
N_n = (1 - Y_e)\,\rho\,/\,m_N ,
\label{eq:vnc}
\end{equation}
where $N_n$ is the neutron number density. For the common $3{+}1$ case, the
mixing matrix is built in the convention
\begin{equation}
\begin{split}
U_{3+1} = {}& R_{34}\,R_{24}(\delta_{24})\,R_{14}(\delta_{14}) \\
          & \times R_{23}\,R_{13}(\delta_{13})\,R_{12} ,
\end{split}
\label{eq:pmns31}
\end{equation}
where $R_{ij}(\delta)$ is a complex rotation in the $(i,j)$ plane and the three
Dirac phases sit on the $(1,3)$, $(1,4)$, and $(2,4)$ rotations. The rotation
ordering alters the physical definitions of $U_{e4}$ and $U_{\mu4}$ and is
stated explicitly to allow direct comparisons with published limits.

\paragraph{Decoherence.} Environmental (Lindblad-type) decoherence and
wave-packet separation both damp interference terms between Hamiltonian
eigenstates,
\begin{equation}
\begin{split}
P_{\alpha\beta} = \sum_{ij} {}& V_{\beta i} V^{*}_{\alpha i} V^{*}_{\beta j} V_{\alpha j} \\
                  & \times e^{-i(w_i - w_j)L}\,D_{ij}(L),
\end{split}
\label{eq:decoh}
\end{equation}
where $V$ and $w$ are the eigenvectors and eigenvalues of the matter Hamiltonian
and $D_{ii}=1$. Two damping models are provided: power-law Lindblad damping
$D_{ij}=\exp(-\Gamma_{ij}L)$ with $\Gamma_{ij}=\gamma_{ij}(E/E_0)^n$, and
Gaussian wave-packet damping $D_{ij}=\exp[-(L/L^{\rm coh}_{ij})^2]$ with
$L^{\rm coh}_{ij} = 2\sqrt{2}\,E\sigma_x / |\Delta w_{ij}|$. In both cases,
$\gamma\to0$ and $\sigma_x\to\infty$ recover standard probabilities exactly.

\paragraph{Non-unitary mixing.} Heavy neutral leptons render the effective
$3\times3$ mixing non-unitary, $N=(1-\alpha)U$, with $\alpha$ lower triangular~\cite{kozynets_nonunitary}.
Propagation is carried out in the mass basis with
\begin{equation}
\begin{gathered}
H_{\rm mass} = \frac{\mathrm{diag}(0,\Delta m^2_{21},\dmsq)}{2E}
              + N^\dagger V_{\rm flavor} N ,\\
V_{\rm flavor} = \mathrm{diag}(V_{\rm CC},0,0) - V_{\rm NC}\mathbb{I},
\end{gathered}
\label{eq:nonunitary}
\end{equation}
where the flavor-universal neutral-current term $V_{\rm NC}\mathbb{I}$ no longer
drops out because $NN^\dagger\neq\mathbb{I}$. The characteristic zero-distance
flavor transition follows,
\begin{equation}
P_{\alpha\beta}(0) = \frac{\left|(NN^\dagger)_{\beta\alpha}\right|^2}
{(NN^\dagger)_{\alpha\alpha}\,(NN^\dagger)_{\beta\beta}} .
\label{eq:zerodistance}
\end{equation}
Antineutrinos take $N\to N^{*}$ alongside $V\to -V$. Non-unitary mixing is
supported across arbitrary density profiles and the layered Earth; decoherence
is currently restricted to vacuum and a single constant-density layer for the
reasons detailed in section~\ref{sec:limits}.

\paragraph{Solar propagation.} A standard-solar-model electron-density
profile~\cite{bs05} is provided alongside the adiabatic MSW composition of a
neutrino produced at radius $r$ inside the Sun:
\begin{equation}
\begin{gathered}
P_{\alpha\beta} = \sum_i |U_{\beta i}|^2\,F_i(r),\\
F_i(r) = \sum_k \left|\langle\nu_i^{\rm vac}|\nu_k^{m}(r)\rangle\right|^2 w_k ,\\
w_k = \left|\langle\nu_k^{m}(r_{\rm emit})|\nu_\alpha\rangle\right|^2 ,
\end{gathered}
\label{eq:solar}
\end{equation}
where $w_k$ are production weights and phase coherence between mass eigenstates
is lost over the Sun--Earth baseline, making the outer sum incoherent.
This reproduces the LMA solution, with $P_{ee}$ falling from the
vacuum-averaged $\sum_i|U_{ei}|^4$ at sub-MeV energies to the matter-dominated
$\cos^4\theta_{13}\sin^2\theta_{12}+\sin^4\theta_{13}$ above $\sim\!10\,$MeV
(where $\theta_{13}$ matter-coupling corrections represent a $4\%$ effect at the NuFIT point).

\section{Numerical core}
\label{sec:numerics}

The primary requirement for an oscillation AD engine is numerical precision across both forward probabilities and reverse-mode derivative sweeps.
Double precision (\texttt{float64}) is necessary rather than merely desirable: oscillation phases involve small mass-squared differences ($\dmsq\sim10^{-3}\,$eV$^2$) accumulated over planetary baselines ($10^4\,$km), so the phase is a small difference of large numbers.
Repeating a core-crossing PREM calculation ($\costheta=-1$, $1$--$20$\,GeV) in single precision shifts probabilities by up to $8\times10^{-6}$ (r.m.s.\ $7\times10^{-7}$).
While modest for forward probability estimates, this single-precision numerical noise is four to eight orders of magnitude above the agreements required for exact derivative evaluation and cross-code benchmarks.
\mango{} therefore enforces 64-bit arithmetic across all computational backends on import.

Evaluating equation~(\ref{eq:product}) requires a unitary matrix exponential for each segment.
\mango{} provides four interchangeable backends for that exponential, each able to stand in for any other inside the layered product.
Across all parameter regimes, including exact spectral degeneracies, \mango{} categorizes its four backends into primary production engines (items 1--2) and reference or extension baselines (items 3--4), agreeing to $10^{-10}$ or better (typically at the $10^{-15}$ level) as continuous mutual cross-checks: \begin{enumerate} \item \textbf{Divided differences} (default production engine).
The matrix function $\exp(-iHL)$ is evaluated directly from the eigenvalues of $H$ via a Cayley--Hamilton expansion using Newton divided differences of $\exp$, avoiding explicit eigenvector computation.
By bypassing explicit eigenvector computation, this backend avoids the derivative singularities ($\propto 1/(\lambda_i-\lambda_j)$) that plague standard diagonalization routines near degenerate spectra, providing $C^\infty$ smooth AD derivatives across exact mass level-crossings.
For three-flavor propagation ($N=3$), eigenvalues are solved in closed form via the trigonometric cubic formula, making the kernel ultra-fast and fully JAX-native.
Because the divided-difference expansion is exact in the confluent limit, supplied with exact eigenvalues it reproduces $\exp(-iHL)$ to machine precision even for an exactly degenerate spectrum.
In practice, the accuracy at degeneracy is set not by the expansion but by the eigenvalue solve: the trigonometric solution of the characteristic cubic passes through $\arccos(r)/3$, whose derivative diverges as $r\to\pm1$, and it therefore resolves an exactly degenerate pair only to $\sim\!5\times10^{-7}$ of the spectral radius.
The resulting propagator error, measured against a \textsc{SciPy}~\cite{scipy} reference over $500$ exactly degenerate Hermitian matrices normalized to unit spectral radius and one radian of accumulated phase, is $5\times10^{-14}$ typical and $9\times10^{-14}$ worst case, which is well within physical requirements while guaranteeing AD safety.
\item \textbf{NuFast port} (constant-density baseline).
A JAX transcription of the NuFast-LBL algorithm~\cite{nufast}, in which the matter mixing-matrix magnitudes and the matter Jarlskog invariant are obtained analytically from the matter eigenvalues and inserted into the standard oscillation formula.
Integrated under its original MIT licence, this provides a fast constant-density path ($61\,$ns on GPU) and an \emph{algorithmic} cross-check that validates the general matrix machinery against closed-form LBL expressions.
\item \textbf{Dense Hermitian diagonalization} (\texttt{jax.linalg.eigh}).
Serves as the general reference path for $3{+}N$ sterile extensions ($N > 3$) where closed-form divided-difference expansions are not yet deployed.
Because its AD rule contains $1/(\lambda_i-\lambda_j)$ terms from eigenvector derivatives, it is treated as a fallback path for non-degenerate sterile configurations (section~\ref{sec:limits}).
\item \textbf{Padé matrix exponential} (\texttt{jax.scipy.linalg.expm}).
Included strictly as a numerical ground-truth baseline to verify custom propagation kernels against standard library routines in continuous integration test suites.
\end{enumerate}

Table~\ref{tab:timing} compares the backends on CPU and on an NVIDIA A100 in double precision.
The NuFast port is the fastest constant-density route, at $61$\,ns per energy point on the GPU, and agrees with the general divided-difference kernel to $10^{-14}$; among the general routes the divided-difference kernel leads at $116$\,ns, ahead of dense Hermitian diagonalization ($253$\,ns) and the dense exponential ($326$\,ns).
The layered PREM path costs three to four orders of magnitude more per point than a single-segment calculation (a factor of $1.4\times10^{3}$ on the GPU and $1.6\times10^{4}$ on the CPU against the divided-difference row), as expected from the ordered product over $\mathcal{O}(50)$ segments.
That is precisely the regime in which a fast constant-density analytic method such as \textsc{NuFast}~\cite{nufast} or \textsc{CHIC}~\cite{chic} does not directly apply, and in which the ability to obtain derivatives without re-deriving them by hand is most valuable.
The GPU buys roughly an order of magnitude on constant-density batches and two orders of magnitude on the layered path, where the many small $3\times3$ operations parallelize well across the batch.
All entry points are \texttt{jit}-compilable and \texttt{vmap}-batchable over energies, zenith angles and parameter sets, and the same source runs on CPU and GPU (both are benchmarked in Table~\ref{tab:timing}).
Nothing in \mango{} is device-specific, so TPU execution should follow.

\begin{table*}[!t]
\caption{Backend performance per energy point, float64, \texttt{jit}-compiled
and batched after warm-up (2000 energy points for constant density, 200 for the
PREM path at $\costheta=-0.8$). CPU: 8 cores of an AMD EPYC 7542; GPU: one
NVIDIA A100. Timings are reproducible to about $15\%$ between runs. Reverse-mode derivative
costs are given separately in table~\ref{tab:gradcost}.}
\label{tab:timing}
\begin{center}
\begin{tabular}{lll}
\toprule
configuration & CPU & A100 \\
\midrule
constant density, NuFast port & 498\,ns & 61\,ns \\
constant density, divided differences & 990\,ns & 116\,ns \\
constant density, dense Hermitian diagonalization & 1.78\,$\mu$s & 253\,ns \\
constant density, dense \texttt{expm} & 8.86\,$\mu$s & 326\,ns \\
PREM Earth & 16.1\,ms & 158\,$\mu$s \\
\bottomrule
\end{tabular}
\end{center}
\end{table*}

\section{Differentiability}
\label{sec:autodiff}

Because every operation in \mango{} is constructed from JAX primitives, automatic differentiation yields exact derivatives of any output with respect to any input in either forward or reverse mode.
Every sector defined in section~\ref{sec:formalism} is fully differentiable, spanning five distinct classes of parameters:

\begin{itemize}
\item \textbf{Oscillation parameters}: $\partial P/\partial\theta_{ij}$,
$\partial P/\partial\delta_{\rm CP}$, and $\partial P/\partial\Delta m^2_{ij}$.
These are the quantities \textsc{CHIC}~\cite{chic} obtains in closed form.
\item \textbf{Geometry}: $\partial P/\partial\costheta$,
$\partial P/\partial h_{\rm atm}$, and $\partial P/\partial d_{\rm det}$. These
flow directly through the chord-segmentation construction of
section~\ref{sec:formalism} and have no counterpart in constant-density analytic
formulas.
\item \textbf{Matter distribution}: $\partial P/\partial\rho_k$ and
$\partial P/\partial Y_{e,k}$ for individual Earth shells, providing the exact
Jacobian required to infer density structure from oscillation data.
\item \textbf{Solar profiles and MSW level-crossings}: $\partial P_{ee}/\partial\theta_{12}$,
$\partial P_{ee}/\partial(\Delta m^2_{21})$, $\partial P_{ee}/\partial\ln N_e(0)$, and
functional core tomography gradients $\delta P_{ee}/\delta\ln N_e(r)$, differentiating
smoothly through continuous solar density profiles and MSW resonance transitions.
\item \textbf{Beyond-Standard-Model sectors}: $\partial P/\partial\varepsilon_{\alpha\beta}$,
$\partial P/\partial\theta_{i4}$, $\partial P/\partial\gamma_{ij}$, and
$\partial P/\partial\alpha_{\alpha\beta}$. These provide exact Jacobians for
limit-setting fits and model constraints without manual derivations.
\end{itemize}

\begin{figure*}[!t]
\includegraphics[width=\textwidth]{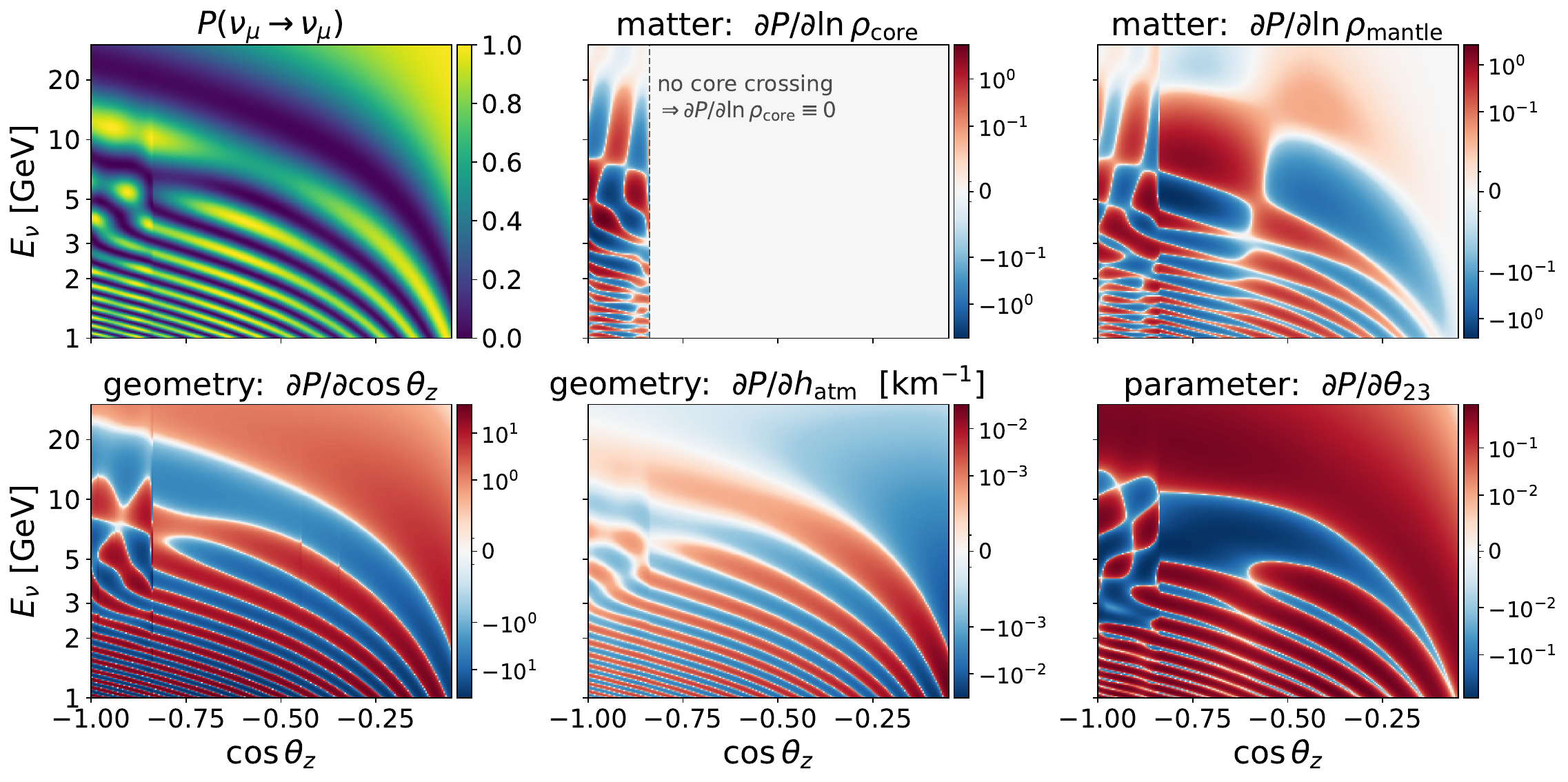}
\caption{Muon-neutrino survival probability through the PREM Earth (top left)
and exact derivatives spanning three standard classes of
section~\ref{sec:autodiff} (two matter, two geometry, one parameter), all
produced by a single \texttt{jacfwd} on the same grid. \emph{Matter}: the
logarithmic core and mantle densities; the core derivative vanishes identically
outside the core-crossing region ($\costheta\lesssim-0.84$), reproducing the
expected geometric threshold. \emph{Geometry}: the zenith angle and the
atmospheric production height. Neither is a parameter of any oscillation formula;
both exist here only because the chord construction is differentiated. Note the
scales: $\partial P/\partial\costheta$ reaches $\pm40$ while $\partial
P/\partial h_{\rm atm}$ is of order $10^{-2}\,$km$^{-1}$ and confined to
near-horizon trajectories, where the atmospheric leg is an appreciable fraction
of the path. \emph{Parameter}: $\theta_{23}$, shown for scale. Each derivative
panel uses a symmetric logarithmic colour scale, linear within $1/100$ of that
panel's own peak; the structure spans two decades, and on a linear scale
everything but the largest excursion washes out. Computed on a $220\times220$
grid, $E_\nu\in[1,30]\,$GeV logarithmically and $-1\le\costheta\le-0.05$, with
$n_{\rm sub}=4$ and $h_{\rm atm}=15\,$km.}
\label{fig:rhograd}
\end{figure*}

\begin{figure*}[!t]
\includegraphics[width=\textwidth]{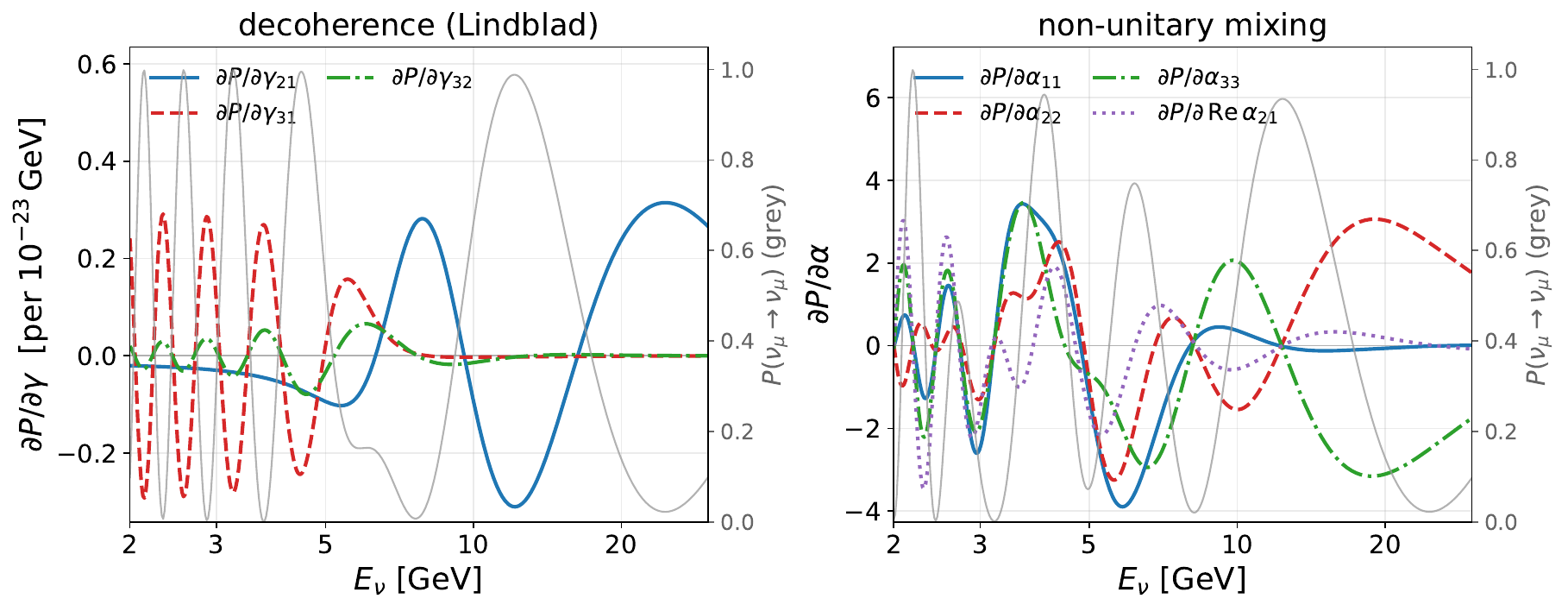}
\caption{Derivatives with respect to the beyond-standard-model parameters
themselves, evaluated at the standard-model point, from the same reverse pass
as any other parameter sensitivity. (a)~Lindblad decoherence, $\partial
P(\nu_\mu\to\nu_\mu)/\partial\gamma_{ij}$ through the Earth's diameter
($12742\,$km) at a constant $5.5\,$g\,cm$^{-3}$, in units of $10^{-23}\,$GeV.
(b)~Non-unitary mixing, $\partial P/\partial\alpha_{ij}$ through the
PREM Earth at $\costheta=-1$; the $\alpha$ are dimensionless. The survival
probability itself is the grey curve on the right-hand axis of each panel. Both
are shown from $2\,$GeV, below which the oscillation phase through a diameter
turns over faster than the figure resolves.}
\label{fig:bsmderiv}
\end{figure*}

\begin{figure*}[!t]
\includegraphics[width=\textwidth]{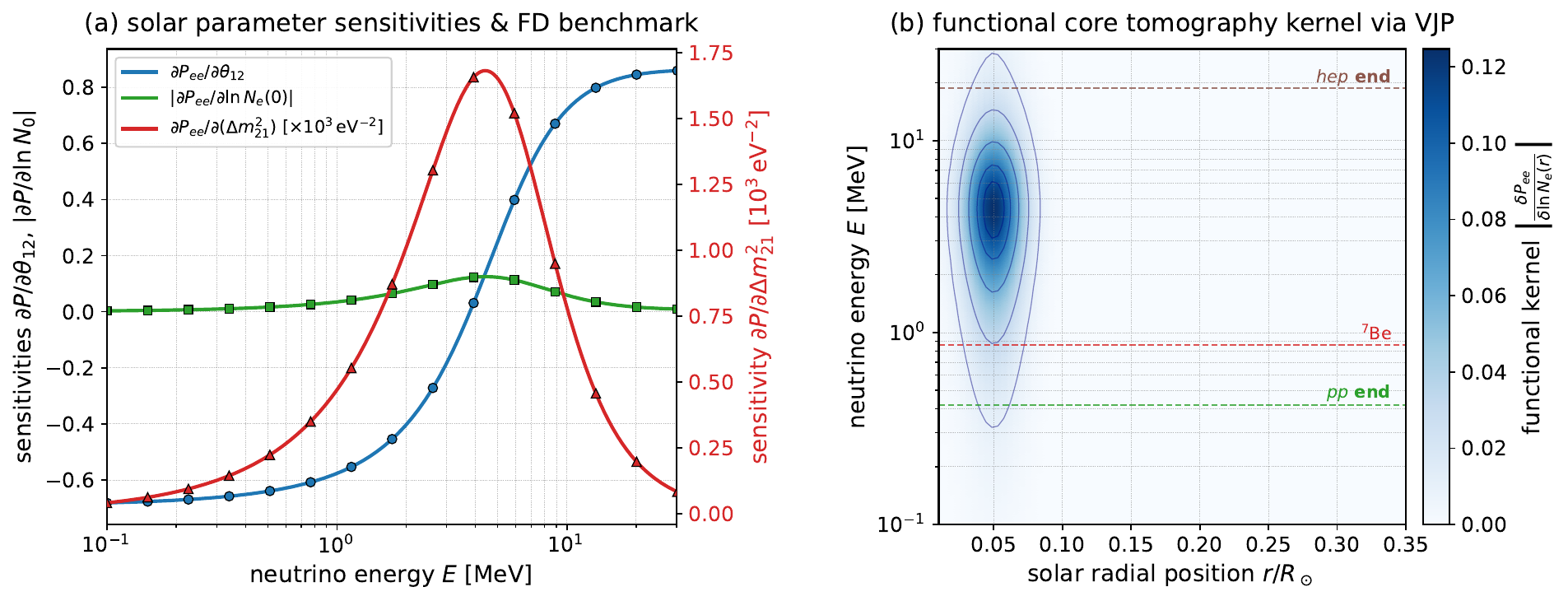}
\caption{Solar neutrino oscillation Automatic Differentiation benchmark and functional core tomography kernel. 
\emph{Panel (a)}: Exact AD parameter sensitivities $\partial P_{ee}/\partial\theta_{12}$ (blue, left axis), $|\partial P_{ee}/\partial\ln N_e(0)|$ (green, left axis), and $\partial P_{ee}/\partial(\Delta m^2_{21})$ (red, right axis) across $E_\nu \in [0.1, 30]\,$MeV, featuring the sharp MSW resonance peak at $E = 4.41\,$MeV ($1.68\times 10^3\,\text{eV}^{-2}$). Overlaid markers show two-point central finite-difference evaluations matching AD values to $1.3\times 10^{-10}$ relative precision. 
\emph{Panel (b)}: 2D functional solar core tomography kernel $|\delta P_{ee} / \delta \ln N_e(r)|$ evaluated across $80$ radial shells $r \in [0.01, 0.35]\,R_\odot$ in a single reverse-mode VJP pass ($1.6\,$s). Horizontal dashed lines indicate key solar neutrino energy thresholds ($pp$ endpoint at $0.42\,$MeV, $^7\text{Be}$ line at $0.86\,$MeV, and $hep$ endpoint at $18.8\,$MeV).}
\label{fig:solarautodiff}
\end{figure*}

Figure~\ref{fig:rhograd} shows representative derivatives spanning the three standard PREM Earth propagation classes defined above (five in total: two matter distribution, two geometry, and one standard oscillation parameter), computed simultaneously by a single \texttt{jacfwd} pass on the same grid. (Derivatives for extended BSM sectors and continuous solar profiles are presented separately in Figures~\ref{fig:bsmderiv} and \ref{fig:solarautodiff}.) Forward mode is used here because evaluating full grid maps requires computing many output points from a small set of differentiated inputs.

The two matter panels carry the clearest physical signature.
The core sensitivity is identically zero for $\costheta\gtrsim-0.84$ and switches on abruptly at the core-crossing threshold, while the mantle derivative extends across the entire up-going sky.
Together, they form the sensitivity map that a tomographic inference of the Earth's interior uses as its Jacobian.
Building this map by finite differences would cost two probability evaluations per shell parameter and remain limited by step-size tuning rather than code precision.

The geometry panels highlight derivatives with no closed-form analytic counterpart.
Standard probability formulas take total baseline $L$ as a single scalar input, whereas in a layered Earth, $\costheta$, $h_{\rm atm}$, and $d_{\rm det}$ non-linearly determine the entire vector of individual shell segment lengths $\{L_k\}$.
These geometric derivatives exist because the trajectory chord construction mapping $(\costheta, h_{\rm atm}, d_{\rm det}) \to \{L_k\}$ is differentiated directly alongside the propagation product. $\partial P/\partial\costheta$ is the largest derivative in magnitude, reaching $\pm40$ on this grid, and inherits both the core-crossing threshold and the dense near-horizon fringes.
In contrast, $\partial P/\partial h_{\rm atm}$ is smaller by three orders of magnitude and is strictly confined to the near-horizon region, correctly reflecting that atmospheric production height matters only where the atmospheric leg constitutes an appreciable fraction of the total path.

While forward mode is ideal for grid-based visualization, reverse mode (\texttt{jax.grad} or \texttt{jax.jacrev}) unlocks the primary computational advantage of automatic differentiation: evaluation cost is governed by the dimension of the output rather than the input.
Table~\ref{tab:gradcost} demonstrates this scaling across standard and geometric derivative classes, timing Jacobians for standard oscillation parameters (6 quantities), trajectory geometry (42 quantities), and the full parameter set of a $123$-shell \texttt{LayeredEarth} (369 parameters spanning densities, electron fractions, and boundary radii).

The matter case provides the most stringent test of this output-dimension scaling.
Unlike the geometry case, where each zenith angle enters only its own trajectory, every shell parameter enters every trajectory that crosses it.
Yet, expanding the input count by a factor of 62 (from 6 to 369 parameters) increases the reverse-mode cost from $2.5\times$ to only $3.0\times$ forward pass evaluations across both CPU and GPU architectures.
Evaluating those same 369 parameters via central finite differencing would demand 738 forward pass evaluations (roughly 250 times the reverse-mode cost) while suffering severe numerical ill-conditioning wherever phase variations are rapid.

\begin{table*}[!t]
\caption{Cost of a reverse-mode AD pass relative to one forward evaluation,
by derivative class, on an NVIDIA A100 in double precision, batched over $40$
energy/zenith points. Each entry is the mean and standard deviation over six
independent groups of calls. The matter row uses the parametric
\texttt{LayeredEarth} representation and is therefore quoted against its own
forward evaluation rather than the default PREM one.}
\label{tab:gradcost}
\begin{center}
\begin{tabular}{lrll}
\toprule
derivative with respect to & parameters & time [ms] & cost / forward \\
\midrule
(forward, default PREM)            &   0 & $1.75\pm0.01$ & $1$ \\
oscillation parameters              &   6 & $4.41\pm0.16$ & $2.5$ \\
geometry ($\costheta$, $h_{\rm atm}$, $d_{\rm det}$) & 42 & $5.16\pm0.08$ & $2.9$ \\
\midrule
(forward, \texttt{LayeredEarth})     & 0 & $4.67\pm0.03$ & $1$ \\
shell densities, $Y_e$ and radii   & 369 & $13.9\pm0.9$  & $3.0$ \\
\bottomrule
\end{tabular}
\end{center}
\end{table*}

The evaluation ratio exhibits minor batch dependence: $2.5$ corresponds to a mixed up- and down-going batch, whereas an all-upgoing batch yields $3.0$ for the six oscillation parameters.
BSM extensions also incur no computational overhead relative to standard propagation because they execute through the same $N\times N$ matrix-exponential kernels (Table~\ref{tab:timing}).

The derivatives that best illustrate what BSM sectors gain from \mango{} are $\partial P/\partial\gamma$ and $\partial P/\partial\alpha_{\beta\alpha}$, the decoherence and non-unitarity parameters themselves.
These are the exact quantities a sensitivity forecast for such models requires, and they exist nowhere in closed form.
Figure~\ref{fig:bsmderiv} displays these BSM sensitivities evaluated at the Standard Model point ($\gamma=0$, $\alpha=0$), where limit-setting fits linearize.
Scaled to $10^{-23}\,$GeV (the order of current atmospheric bounds on Lindblad decoherence), $\partial P/\partial\gamma$ reaches $0.3$, while the dimensionless $\alpha$ derivatives reach $3$--$4$ per unit $\alpha$.
Automatic differentiation evaluates these non-analytic BSM sensitivities directly at the Standard Model linearization point where limit-setting fits execute.
As shown in Figure~\ref{fig:bsmderiv}, reverse-mode AD captures the detailed energy dependence and parameter separation across the $\theta_{13}$ MSW resonance (e.g., distinguishing $\partial P/\partial\gamma_{31}$ from $\partial P/\partial\gamma_{21}$) in a single reverse pass, eliminating the need for manual symbolic derivations or expensive multi-parameter finite-difference scans.

In addition to step-segmented planetary profiles, \mango{} supports continuous $C^\infty$ matter distributions, enabling functional sensitivity analysis of solar neutrino propagation.
Figure~\ref{fig:solarautodiff}(a) shows the exact parameter derivatives $\partial P_{ee}/\partial\theta_{12}$, $\partial P_{ee}/\partial(\Delta m^2_{21})$, and $|\partial P_{ee}/\partial\ln N_e(0)|$ across the solar LMA transition region ($0.1$--$30\,$MeV).
The mass-squared splitting sensitivity $\partial P_{ee}/\partial(\Delta m^2_{21})$ exhibits a prominent MSW resonance peak reaching $1.68\times 10^3\,\text{eV}^{-2}$ at $E = 4.41\,$MeV, demonstrating AD's ability to differentiate smoothly through MSW level-crossings without numerical instability.

Beyond scalar parameter derivatives, reverse-mode automatic differentiation enables functional tomography of continuous matter profiles.
Figure~\ref{fig:solarautodiff}(b) displays the 2D functional solar core tomography kernel $|\delta P_{ee}/\delta\ln N_e(r)|$, mapping the sensitivity of survival probability to localized density perturbations across $80$ radial shells ($0.01$--$0.35\,R_\odot$).
Because reverse-mode vector-Jacobian products (VJPs) scale with the output dimension ($O(1)$ in output size), \mango{} computes this entire $80$-shell functional sensitivity kernel across $180$ energy bins in a single VJP pass ($1.6\,$s execution time), whereas a finite-difference construction would require $160$ forward propagation evaluations per energy point.

Because automatic differentiation operates directly on the computational graph, derivatives remain exact by construction and cannot drift out of sync with forward evaluations.
Section~\ref{sec:validation} presents a quantitative validation of all forward models and AD derivatives against analytic limits, independent codes, and finite differences.

\section{Validation}
\label{sec:validation}

\mango{} is validated at four levels: internal consistency (unitarity, $\sum_\beta P_{\alpha\beta}=1$; cross-backend agreement; chord-length geometry against $2R_\oplus|\costheta|$), analytic limits (two- and three-flavor vacuum formulae, the MSW resonance condition), cross-code comparison, and finite-difference checks of every differentiation path.
Table~\ref{tab:validation} collects all of these, separating comparisons against external implementations from limits checked internally, all of which are shipped as automated unit tests.

\begin{table*}[!tb]
\caption{Validation summary. ``cross-code'' rows compare against an external
implementation and give the maximum absolute deviation in probability over the
benchmark set; the \textsc{OscProb} identical-path row feeds \textsc{OscProb}'s
own PREM segmentation into the \mango{} propagator and so isolates the
propagation algorithm, while the remaining Earth comparisons use independently
constructed paths and are dominated by segmentation and electron-fraction
conventions. ``internal limit'' rows have no external reference and instead
check a case whose answer is known exactly (switching an extension off, or a
textbook analytic limit), together with the finite-difference checks of the
corresponding derivatives. All are shipped as unit tests, though the
cross-code rows compare against externally generated reference values frozen
at test time rather than re-run in continuous integration
(section~\ref{sec:usage}).
$^{\P}$\,The \textsc{NuFast} algorithm is formulated for a single
constant-density layer; the companion \textsc{NuFast-Earth} tool applies it to
a layered PREM Earth, and is what the converged-PREM row compares against.
$^{\S}$\,Comparing \mango{}'s three-flavor adiabatic calculation against an
independent three-flavor numerical MSW reference yields machine-precision
agreement ($6\times10^{-15}$). Comparing against the textbook two-flavor
closed-form approximation shows a deviation ($1.3\times10^{-5}$ at $0.1\,$MeV
rising to $1.1\times10^{-3}$ at $15\,$MeV) caused by the reference's
truncation of the $\theta_{13}$ matter-coupling term $\sim\sin^2\theta_{13}\,(2EV_{\rm CC}/\dmsq)$. See section~\ref{sec:validation}.}
\label{tab:validation}
\begin{center}\small
\begin{tabular}{llll}
\toprule
type & check & configuration & deviation \\
\midrule
cross-code & \textsc{OscProb}~\cite{oscprob} & constant density, 1300\,km, $\nu/\bar\nu$ & $6\times10^{-9}$ \\
cross-code & \textsc{OscProb}~\cite{oscprob} & identical PREM path, $\costheta=-1$ & $9\times10^{-8}$ \\
cross-code & \textsc{NuFast}~\cite{nufast} & long-baseline constant density & $2\times10^{-5}$ \\
cross-code & \textsc{NuFast-Earth}~\cite{nufastearth}$^{\P}$ & converged PREM, $\costheta=-1$ & $3\times10^{-6}$ \\
cross-code & 3-flavor numerical MSW$^{\S}$ & adiabatic solar, $0.1$--$15\,$MeV & $6\times10^{-15}$ \\
\midrule
internal limit & NSI & $\varepsilon\to0$ recovers standard & $7\times10^{-15}$ \\
internal limit & $3{+}1$ sterile & $\theta_{i4}\to0$ recovers three-flavor block & $3\times10^{-15}$ \\
internal limit & $3{+}1$ sterile & vacuum $P_{\bar e\bar e}$ vs.\ $1-\sin^2 2\theta_{14}$ & $2\times10^{-7}$ \\
internal limit & $3{+}1$ sterile & unitarity in PREM matter & $8\times10^{-14}$ \\
internal limit & decoherence & Lindblad, $\gamma\to0$ recovers standard & $4\times10^{-15}$ \\
internal limit & decoherence & wave packet, $\sigma_x\to\infty$ recovers standard & $4\times10^{-15}$ \\
internal limit & decoherence & strongly damped, probability conserved & $2\times10^{-15}$ \\
internal limit & non-unitarity & $\alpha\to0$ recovers standard & $2\times10^{-15}$ \\
\midrule
AD vs.\ FD & NSI & $\partial P/\partial\varepsilon_{ee}$, constant density & $3\times10^{-14}$ \\
AD vs.\ FD & $3{+}1$ sterile & $\partial P/\partial\theta_{14}$, PREM & $2\times10^{-10}$ \\
AD vs.\ FD & $3{+}1$ sterile & $\partial P/\partial\Delta m^2_{41}$ \emph{at} $\Delta m^2_{41}=\dmsq$ & $12\times$ (see text) \\
\bottomrule
\end{tabular}
\end{center}
\end{table*}

When comparing \mango{} against external Earth propagation tools (such as \textsc{NuFast-Earth} in Table~\ref{tab:validation}), a small residual difference of $\mathcal{O}(10^{-6})$ remains.
To verify that this discrepancy is not caused by numerical discretization error within \mango{}, we perform a convergence test by refining the sub-shell segmentation parameter $n_{\rm sub}$ (the number of sub-steps per PREM layer).
As $n_{\rm sub}$ increases from $5$ to $800$, the deviation drops from $3.2\times10^{-5}$ and plateaus at $3.07\times10^{-6}$ (at $2\,$GeV, and $7.4\times10^{-6}\to9.9\times10^{-7}$ at $5\,$GeV), achieving full numerical convergence by $n_{\rm sub}\approx 100$.
This confirms that \mango{}'s step-segmented integration is completely converged, proving that the tiny remaining residual stems entirely from minor convention differences in electron-fraction profiles ($Y_e$) and shell boundary radii between independent implementations.

Because evaluating exact derivatives is the core purpose of \mango{}, every differentiation path is verified against central finite differences of the forward model.
Table~\ref{tab:adfd} reports the results across all derivative classes defined in section~\ref{sec:autodiff}, evaluated over a grid of energies and zenith angles spanning core- and mantle-crossing trajectories.
The agreement for standard parameter and density derivatives ranges from $3\times10^{-10}$ ($\theta_{13}$) to $6\times10^{-9}$ ($\delta_{\rm CP}$), bounded by finite-difference truncation error rather than automatic differentiation accuracy, as the two calculations share no code beyond the forward model.
Solar neutrino derivatives through continuous matter profiles match central finite-difference evaluations to $1.3\times 10^{-10}$ relative precision (Figure~\ref{fig:solarautodiff}(a)).
The $\partial P/\partial\costheta$ row is the single exception at $7\times10^{-7}$ due to the nadir endpoint ($\costheta=-1$), where central differences are unavailable and a one-sided substitute is used (see Table~\ref{tab:adfd} caption).
This benchmark confirms the accuracy of geometric derivatives, which must correctly account for the shifting of shell boundary crossings along the trajectory chord.

\begin{table*}[!tb]
\caption{Automatic differentiation against central finite differences of the
same forward model, maximum over a $24$-point grid spanning
$1$--$20$\,GeV and $-1\le\costheta\le-0.2$, \emph{including the nadir endpoint
$\costheta=-1$}. Central differences cannot step past $\costheta=-1$, so that
one point uses a second-order one-sided difference; its truncation error is
what sets the $\partial P/\partial\costheta$ row, and the one-sided estimate
converges onto the automatic-differentiation value at the expected
second-order rate as the step is reduced, by relative
$8\times10^{-4}$, $3\times10^{-6}$, $3\times10^{-8}$ and $7\times10^{-10}$ for
$h=10^{-4}\ldots10^{-7}$. The last column is the largest
absolute value of the derivative over that grid, in probability per unit of the
quantity differentiated (radians, eV$^2$, km, or a dimensionless density
scaling), and sets the scale against which the deviation should be read.}
\label{tab:adfd}
\begin{center}
\begin{tabular}{llll}
\toprule
derivative & class & max.\ rel.\ deviation & $|\partial P|_{\max}$ \\
\midrule
$\partial P/\partial\theta_{13}$          & parameters & $3\times10^{-10}$ & $6.5$ \\
$\partial P/\partial\theta_{23}$          & parameters & $2\times10^{-9}$  & $0.82$ \\
$\partial P/\partial\delta_{\rm CP}$      & parameters & $6\times10^{-9}$  & $1.3\times10^{-2}$ \\
$\partial P/\partial\dmsq$                & parameters & $4\times10^{-10}$ & $1.6\times10^{4}$ \\
$\partial P/\partial\costheta$            & geometry   & $7\times10^{-7}$  & $37$ \\
$\partial P/\partial h_{\rm atm}$         & geometry   & $4\times10^{-10}$ & $3.0\times10^{-3}$ \\
$\partial P/\partial\ln\rho_{\rm core}$   & matter     & $5\times10^{-10}$ & $2.1$ \\
$\partial P/\partial\ln\rho_{\rm mantle}$ & matter     & $7\times10^{-10}$ & $1.7$ \\
\bottomrule
\end{tabular}
\end{center}
\end{table*}

The beyond-standard-model sectors introduced in section~\ref{sec:formalism} (NSI, sterile states, decoherence, and non-unitary mixing) are validated through exact physical limits where closed-form answers are known independently (Table~\ref{tab:validation}): disabling an extension recovers standard three-flavor propagation bit-for-bit through an independent code path, the $3{+}1$ vacuum survival probability reaches $1-\sin^2 2\theta_{14}$ at the $\Delta m^2_{41}$ oscillation maximum, and probability is conserved in matter.
Sensitivities for all BSM parameters are verified against central finite differences on the same footing as standard parameters.

Finally, the atmospheric oscillograms of figure~\ref{fig:oscillogram} provide a qualitative verification of key physical features, including the MSW resonance, the $\nu/\bar\nu$ matter asymmetry, the core--mantle boundary at $\costheta\simeq-0.84$, and dense near-horizon oscillation fringes.

\section{A worked example: tomographic sensitivity end to end}
\label{sec:worked}

To demonstrate how these derivatives, whether standard or BSM, are propagated into downstream analysis metrics, we carry the derivatives through to a quantity that is not a probability at all: the uncertainty with which a counting experiment could constrain the density of the Earth's core, and how that precision responds to detector design parameters.

The primary tool is the Fisher information matrix $F$, whose inverse $F^{-1}$ yields the parameter covariance matrix in the Gaussian approximation; $\sqrt{(F^{-1})_{ii}}$ is therefore the uncertainty on parameter $i$ after marginalizing over all other free parameters.
Core density is measured not in isolation but in direct competition with oscillation parameters, atmospheric flux nuisances, and adjacent density shells.
Constructing $F$ requires expected count derivatives with respect to every parameter: \begin{equation} F_{ij}=\sum_b \mu_b^{-1}\, \frac{\partial\mu_b}{\partial\vartheta_i}\frac{\partial\mu_b}{\partial\vartheta_j}, \label{eq:fisher} \end{equation} evaluated over a toy atmospheric sample of $10^5$ $\nu_\mu$ events (plus $0.5\times 10^5$ $\bar\nu_\mu$ events) binned over $E\in[1,30]\,$GeV and $\costheta\in[-1,-0.05]$ ($30\times 30$ log-energy and linear-zenith bins).
The atmospheric flux follows $\Phi\propto(E/10\,\mathrm{GeV})^{-(2.7+\Delta\gamma)}$ at an atmospheric production height of $h_{\rm atm}=15\,$km, while the baseline detector response applies a $20\%$ fractional energy resolution ($\sigma_{\ln E}=0.20$) and an angular resolution operating point of $\sigma_{\costheta}=0.10$.

\begin{figure*}[!t]
\includegraphics[width=\textwidth]{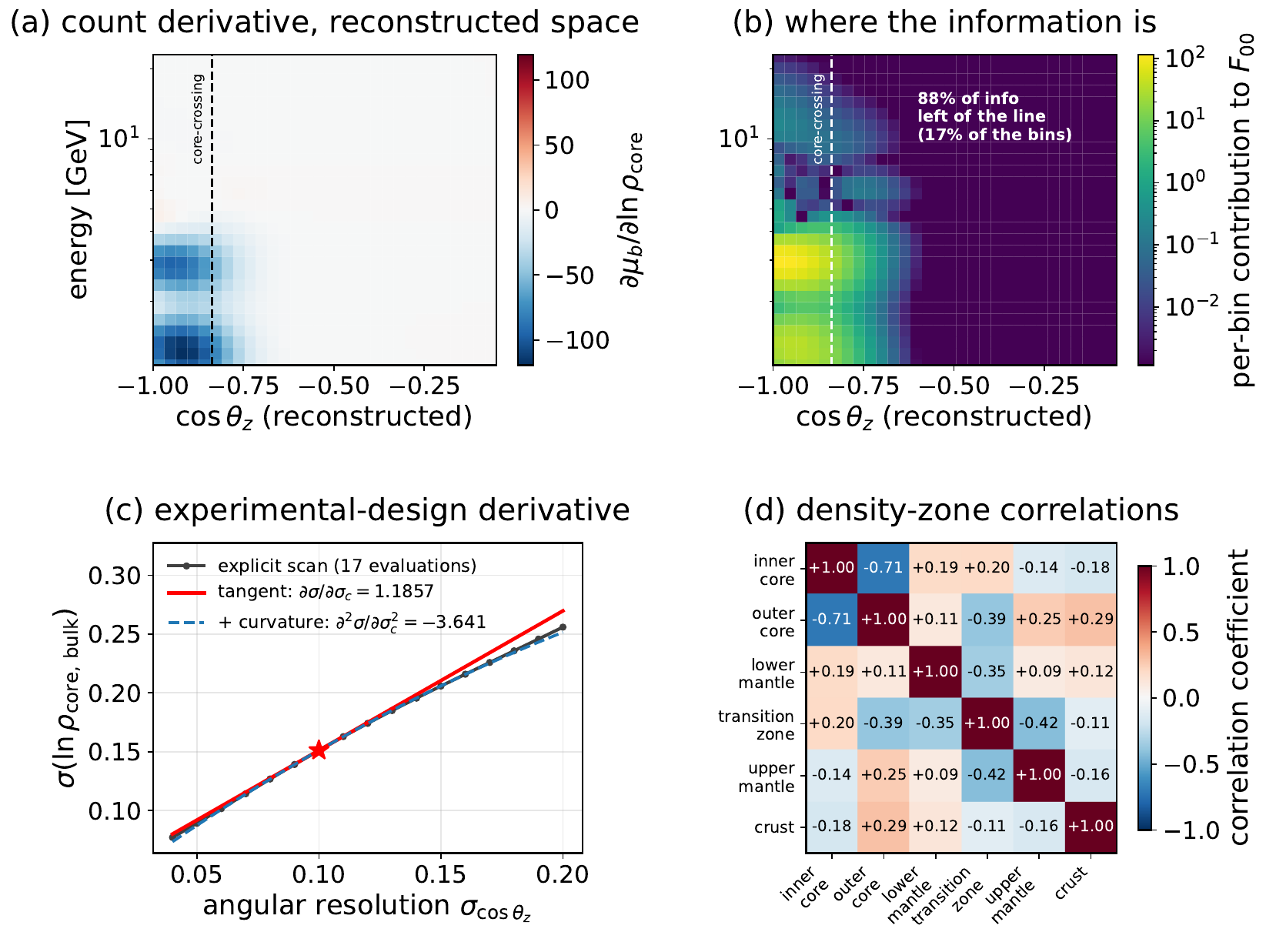}
\caption{The differentiation chain of section~\ref{sec:worked}. (a) Derivative
of the expected bin counts with respect to the bulk-core density, in
\emph{reconstructed} $(E,\costheta)$ space. (b) The resulting per-bin
contribution to the nuisance-fixed Fisher information, on a logarithmic colour
scale; the dashed line is the core-crossing threshold. (c) The marginalized
$\sigma(\ln\rho_{\rm core,\,bulk})$ against the assumed angular resolution: an
explicit scan (grey, seventeen Fisher evaluations), the tangent from one
automatic-differentiation call at the operating point (red, star), and the
quadratic adding one further nested call (blue, dashed). The tangent is drawn
across the whole range rather than near the operating point, so that where it
ceases to be a good extrapolation is visible. (d) Correlations among the six radial
density parameters; the inner/outer-core pair, outlined, is anticorrelated at
$-0.71$. The colour scale in (a) is symmetric while the derivative is almost
entirely negative, so the positive lobe is faint by construction. See the text
for discussion.}
\label{fig:tomo}
\end{figure*}

The parameter vector $\vartheta$ has eleven entries: six \emph{multiplicative scale factors} on the PREM density, one per radial zone (inner core, outer core, lower mantle, transition zone, upper mantle, crust); $\theta_{23}$ and $\dmsq$; and three flux nuisances (independent $\nu_\mu$ and $\bar\nu_\mu$ normalizations, and a shared spectral tilt $\Delta\gamma$).
Every derivative in equation~(\ref{eq:fisher}) comes from a single \texttt{jacfwd} pass through the complete execution graph: shell densities $\to$ chord geometry $\to$ oscillation probability $\to$ flux weighting $\to$ histogram binning $\to$ detector migration.
For standard workflows, \texttt{mango.stat} automates this binned-Poisson Fisher matrix construction for any model mapping a parameter PyTree to expected counts (with optional Gaussian priors), exposing both the marginalized uncertainty $\sqrt{(F^{-1})_{ii}}$ and the unmarginalized sensitivity $1/\sqrt{F_{ii}}$.
Because \texttt{jacfwd} evaluates derivatives per bin throughout the entire execution graph, \mango{} allows the Fisher information to be traced back to individual reconstructed $(E, \costheta)$ bins (Figure~\ref{fig:tomo}a--b).

Resolving the density into radial zones rather than a single core and mantle factor is what makes the result meaningful.
With six parameters the shells can compete, and the competition is severe: the inner and outer core are anticorrelated at $-0.71$ (figure~\ref{fig:tomo}d), since a denser inner core and a thinner outer core produce nearly the same effect on the observed rates.
For the bulk core (the two together, the region a single $\ln\rho_{\rm core}$ would describe), the uncertainty is $0.021$ with everything else fixed and $\mathbf{0.151}$ once the remaining ten parameters are marginalized over, a factor of seven.\footnote{Both numbers describe the same quantity, the common-mode scaling $q$ in $\vartheta=\vartheta_0+q\,v$ with $v$ the indicator of the two core zones.
The marginalized estimator of a common shift is the average of the per-zone shifts and so carries a factor $1/2$, which must not be dropped in the comparison.} Taking the inner core alone: $0.205$ fixed, $0.255$ marginalizing over the other density zones, $0.210$ over $\theta_{23}$ and $\dmsq$, and $0.334$ over everything.
Competition between density zones therefore costs more than the oscillation parameters do, the opposite of what the two-parameter version of this exercise suggests, where no such competition exists.
Sensitivity scales as $N^{-1/2}$, $F$ being linear in the exposure.

The sensitivity is almost entirely geometric, for a simple reason: a neutrino carries information about the core only if it went through the core.
Below the core-crossing threshold $\costheta=-0.838$ the trajectory passes through the core and the probability responds to its density; above it, the response is exactly zero.
The $17\%$ of bins below the threshold hold $88\%$ of the information on the bulk-core direction while containing only $15\%$ of the events, so the great majority of the data contributes essentially nothing to this particular measurement.
The remaining $12\%$ is not physics but the finite angular resolution, smearing genuinely core-crossing events into bins reconstructed above the threshold.

Figure~\ref{fig:tomo} shows the chain.
Panel (a) is the derivative of the expected \emph{counts}, in reconstructed coordinates, after the flux weighting, the binning and the detector response have all been applied; set against figure~\ref{fig:rhograd}, the derivative of the \emph{probability} in true coordinates, the two differ by exactly the part of the chain that an analytic probability calculator cannot differentiate.
Panel (b) turns that into information and makes the geometric statement above visible.

Panel (c) is the payoff: the sensitivity itself remains differentiable.
Treating the angular resolution as an input and differentiating $\sigma(\ln\rho_{\rm core})$ with respect to it, through the Fisher matrix \emph{and its inverse}, gives \begin{equation} \frac{\partial\,\sigma(\ln\rho_{\rm core,\,bulk})}{\partial\,\sigma_{\costheta}} = 1.186 , \end{equation} reproduced to $1.3\times 10^{-5}$ relative precision against central finite differences, matching the explicit grid scan in Figure~\ref{fig:tomo}(c) to $0.1\%$.
Physically, sharpening the angular resolution by $0.01$ in $\costheta$ improves the core-density precision by $7.8\%$.
Evaluating this experimental-design derivative via AD requires only a single pass through the Fisher matrix inverse, avoiding expensive re-evaluations of the entire analysis pipeline per design parameter.

The resolution enters $\sigma$ only through the Fisher matrix, which is itself built from $\partial\mu/\partial \vartheta$; so $\partial\sigma/\partial\sigma_{\costheta}$ needs the mixed second derivative $\partial^2\mu/\partial\vartheta\,\partial\sigma_{\costheta}$, and is obtained by nesting automatic differentiation: a \texttt{grad} of a \texttt{jacfwd}.
Nothing stops the nesting there.
One more level gives the curvature of the sensitivity, \begin{equation} \frac{\partial^2\,\sigma(\ln\rho_{\rm core,\,bulk})} {\partial\,\sigma_{\costheta}^{\,2}} = -3.64 , \end{equation} which is what says how far the tangent may be extrapolated: over the $0.01$ quoted above the linear estimate is good to $0.1\%$, at $0.02$ to $0.5\%$, and by $0.05$ it is $3\%$ adrift, where the quadratic is close to $2\%$ (figure~\ref{fig:tomo}c).
It is close to free: the inner Jacobian is already an eleven-way forward-mode sweep, so one further tangent direction adds about a tenth of the work.
It agrees with a finite difference of the exact first derivative to $5\times10^{-4}$.
Second derivatives with respect to the \emph{geometry} inherit the shell-grazing cusps of section~\ref{sec:limits}, but the design parameter here is smooth.
Panel (d), finally, presents what the simplified two-parameter version of this exercise could not have produced at all: the complete 11-parameter correlation matrix behind the $-0.71$ inner/outer core anticorrelation noted above.

\section{Scope and future extensions}
\label{sec:limits}

This section outlines the design boundaries of \mango{}'s propagation model alongside key directions for future extensions.

\subsection*{Physical and Geometric Scope}

\paragraph{Unitary amplitude propagation vs. transport.}
\mango{} is designed as a fast, exact, differentiable unitary amplitude propagator for neutrino oscillations. Physical processes that alter total neutrino number (such as high-energy attenuation, absorption, tau-neutrino regeneration, or decay) require a transport equation treatment rather than Hamiltonian evolution and are natively addressed by tools such as \textsc{nuSQuIDS}~\cite{nusquids}.

\paragraph{Varying-density profiles for extended sectors.}
Non-unitary mixing is fully supported across arbitrary continuous profiles and the layered Earth, as the non-unitary operator acts at production and detection around standard unitary evolution. Environmental decoherence is currently available for vacuum and constant-density media. Extending decoherence to varying-density media requires propagating full density matrices and establishing a convention for the damping eigenbasis across density discontinuities, which we reserve for dedicated open-quantum-system modeling.

\paragraph{3D Earth models.}
The standard geometry module models spherically symmetric Earth profiles with differentiable radial densities, electron fractions, and shell radii. Because the propagation core assumes nothing about radial symmetry, extending the front-end chord construction to accommodate 3D lateral heterogeneities requires writing a new trajectory path module without modifying the underlying physics engine.

\subsection*{Numerical and Differentiability Considerations}

\paragraph{Degenerate spectra and backend selection.}
Differentiating explicit eigenvector routines (\texttt{jax.linalg.eigh}) through exact spectral degeneracies introduces numerical derivative singularities due to $1/(\lambda_i-\lambda_j)$ terms. For instance, forcing an exact vacuum mass-level crossing ($\Delta m^2_{41}=\dmsq$) in a $3{+}1$ sterile model leaves forward probabilities unaffected but yields an unreliable AD derivative of $-67.8$ compared to the true finite-difference value of $-5.54$. \mango{} resolves this for standard 3-flavor oscillations by providing the divided-difference backend as its default production engine, which bypasses eigenvector differentiation entirely and guarantees exact, smooth derivatives across degenerate points. For $N > 3$ sterile models, where \mango{} currently falls back to \texttt{jax.linalg.eigh}, extending divided-difference matrix functions to arbitrary $N$-flavor Hamiltonians is a high-priority numerical item scheduled for an upcoming release.

\paragraph{Geometric derivatives at shell boundaries.}
Where trajectory chords graze shell boundaries, entering segment lengths acquire square-root cusps ($\propto\sqrt{r_b^2-r_{\min}^2}$), producing local divergences in $\partial P/\partial\costheta$ over narrow zenith windows ($\sim\!10^{-5}$). This is an inherent geometric property of piecewise-constant shell models rather than an artifact of the implementation. To identify and isolate these geometric non-differentiabilities during sampling or optimization, \mango{} exposes \texttt{earth.critical\_cos\_zenith()}; in practical optimization or sampling workflows, these measure-zero boundary cusps are readily handled via standard gradient clipping, adaptive step-size controls, or smooth boundary softening.

\paragraph{Differentiable inference and experiment integration.}
While \mango{} provides utility parsers to convert GLoBES-style experiment configurations into differentiable JAX execution graphs, its broader role in the computational ecosystem is to enable end-to-end differentiable inference. Because \mango{} registers its parameter structures as native JAX PyTrees, it connects seamlessly with modern probabilistic programming frameworks such as \textsc{NumPyro} and \textsc{BlackJAX}, opening the door to gradient-based Hamiltonian Monte Carlo (HMC/NUTS) sampling, variational inference, and automated experimental design for future global fits.

\section{Code availability and reproducibility}
\label{sec:usage}

\mango{} is a lightweight open-source package of about $3\,000$ lines of pure Python, dependent only on \textsc{JAX}~\cite{jax} and \textsc{NumPy}~\cite{numpy} (with optional \textsc{diffrax}~\cite{diffrax} and \textsc{matplotlib}~\cite{matplotlib} extras).
It is released under the MIT licence and can be installed via \texttt{pip install mango-osc}.
The source code, continuous-integration test suite, and issue tracker are hosted at \url{https://github.com/pgranger23/mango-osc}.
To ensure long-term archival reproducibility, release \texttt{v0.2.4} (from which all benchmarks and figures in this manuscript are derived) is archived on Zenodo under concept DOI \verb|10.5281/zenodo.21774110|.

All validation checks, performance benchmarks, timing measurements, and BSM capability demonstrations reported in this work are fully reproducible via self-contained Python scripts shipped in the \texttt{benchmarks/} and \texttt{examples/} directories.
Every benchmark is configured to run on standard CPU hardware with documented execution times, requiring no GPU acceleration.
The cross-code comparisons of Table~\ref{tab:validation} compare against reference outputs frozen into the regression test suite.

\section{Summary}
\label{sec:summary}

\mango{} provides a foundational shift for neutrino oscillation workflows: moving from traditional forward-only calculators to an exact, end-to-end differentiable engine.
Built natively in JAX, \mango{} evaluates exact derivatives across standard $3$-flavor oscillations, continuous solar matter profiles, PREM-layered Earth geometries, and extended Beyond the Standard Model (BSM) sectors, including matter non-standard interactions (NSI), $3{+}N$ sterile states, environmental decoherence, and non-unitary mixing.

The package addresses the key numerical and computational challenges of differentiating quantum evolution: \begin{enumerate} \item \textbf{Singularity-Free Differentiation}: By adopting matrix divided-difference expansions as its primary production backend, \mango{} bypasses explicit eigenvector differentiation, eliminating numerical derivative singularities ($1/(\lambda_i - \lambda_j)$) across spectral degeneracies and MSW level-crossing resonances.
\item \textbf{Sub-linear AD Scaling}: Reverse-mode automatic differentiation evaluates complete Jacobian matrices (encompassing standard oscillation parameters, complex BSM couplings, and $369$ PREM Earth density, electron-fraction, and shell-radius parameters) at a minor constant overhead ($2.5$--$3\times$) relative to a single forward evaluation.
\item \textbf{Rigorous Validation}: Every differentiation path is validated against central finite differences, matching forward probabilities to $10^{-9}$--$10^{-5}$ against established tools (\textsc{OscProb}, \textsc{NuFast}, \textsc{Nuwaves}) and derivative evaluations to $10^{-10}$--$10^{-5}$ across all physical limits.
\end{enumerate}

By preserving exact sensitivities from upstream Hamiltonian evolution through geometry, flux weighting, binning, and detector migration, \mango{} enables three main classes of downstream physics applications: \begin{itemize} \item \textbf{Experimental Design and Binning Optimization}: Differentiating expected counts with respect to histogram bin edges ($(E_b, \cos\theta_{z,b})$) and detector resolution parameters allows algorithmically optimizing binning schemes for maximum Fisher information, while evaluating how hardware upgrades trade off between Earth tomography and new-physics discovery reach in a single AD pass.
\item \textbf{Differentiable Inference and High-Dimensional Fits}: Registering parameter structures as native JAX PyTrees connects \mango{} directly to modern gradient-based minimizers and probabilistic frameworks (\textsc{NumPyro}, \textsc{BlackJAX}), enabling fast marginalization over hundreds of systematic nuisance parameters and efficient Hamiltonian Monte Carlo (HMC/NUTS) sampling.
\item \textbf{Geophysical Inversions and Functional Tomography}: Leveraging exact functional sensitivity kernels ($\delta P/\delta\rho(r)$) enables gradient-based inversion techniques to reconstruct continuous 1D/3D Earth interior density and electron-fraction profiles directly from atmospheric neutrino data.
\end{itemize}

As neutrino physics transitions toward differentiable data analysis pipelines, \mango{} provides the open-source, high-performance foundation connecting fundamental quantum mechanics directly to experimental discovery potential.

\section*{Acknowledgements}
The author thanks the developers of \textsc{OscProb} and \textsc{Nuwaves}, whose reference
implementations were used throughout validation and cross-checks. \mango{} redistributes
a JAX transcription of the \textsc{NuFast-LBL} reference implementation of
Denton and Parke under its MIT licence, with the original copyright retained;
the author is grateful to them for releasing it openly.

\section*{Data availability statement}
The source code, test suite and benchmark data that support the findings of
this study are openly available at \url{https://github.com/pgranger23/mango-osc}
and archived on Zenodo. The results reported here were produced with release
\texttt{v0.2.4}. The concept DOI \verb|10.5281/zenodo.21774110| resolves to
the archived release.

\end{document}